\documentclass[
aps,%
final,%
notitlepage,%
oneside,%
twocolumn,%
nobibnotes,%
nofootinbib,% В текущей версии REVTeX c этой опцией
             не работают сноски (footnotes) в таблицах.
superscriptaddress,%
noshowpacs,%
centertags]%
{revtex4}
\usepackage[dvipsone]{graphicx}
\usepackage[english]{babel}
\begin{document}
%\selectlanguage{russian} % Для  статьи на русском языке
%\selectlanguage{english} % Для статьи на английском языке
%
\title{Model of the tail region of the heliospheric interface}
\author{\firstname{Vlad} \surname{Izmodenov}}
\email[]{izmod@ipmnet.ru}
\homepage[]{http://izmod.ipmnet.ru/~izmod/}
%\thanks{}
\affiliation{Lomonosov Moscow State University, Moscow, Russia}
%}%Институт Проблем Механики РАН}
\author{\firstname{Dmitry} \surname{Alexashov}}
%\author{\firstname{Дмитрий} \surname{Алексашов}}
\email[]{alexash@ipmnet.ru}
%\thanks{}
%\altaffiliation{}
\affiliation{Institute for Problems in Mechanics Russian Academy
of Sciences}
%
%\noaffiliation
%
\begin{abstract}
Physical processes in the tail of the solar wind interaction
region with the partially ionized local interstellar medium are
investigated in a framework of the self-consistent kinetic-gas
dynamic model. It is shown that the charge exchange process of the
hydrogen atoms with the plasma protons results in suppression of
the gas dynamic instabilities and disappearance the contact
discontinuity at sufficiently (~3000 AU) large distances from the
Sun. The solar wind plasma temperature decreases and, ultimately,
the parameters of the plasma and hydrogen atoms approach to the
corresponding parameters of the unperturbed interstellar medium at
large heliocentric distances.

\end{abstract}

\maketitle

\newpage
\section{Introduction}

The Sun and its solar system are presently moving in the partly
ionized local interstellar cloud (LIC) (e.g., Lallement, 1996).
Direct measurements of interstellar atoms of helium (Witte et al.,
1996) by GAS/Ulysses experiment show that the velocity of relative
Sun-LIC motion is about 25 km/sec, and local interstellar
temperature is about 6000 K. Other interstellar parameters, such
as interstellar ionization state, densities of neutral and charged
components, magnitude and direction of the interstellar magnetic
field, can be determined by remote space experiments. These are e
measurements of backscattered solar Lyman $\alpha$ radiation on
boards of SOHO, Voyager, Pioneer spacecraft,  pickup ions on
boards of Ulysses and ACE spacecraft, the solar wind properties at
large heliocentric distances by Voyager, absorption of Lyman
$\alpha$ spectra toward nearby stars, and heliospheric fluxes of
energetic neutral atoms (ENA). An adequate theoretical model of
the solar wind interaction with LIC is needed to interpret the
remote experiments. Theoretical concept of the solar wind
interaction with the solar wind was proposed in the pioneer paper
by Baranov et al. (1970). During last 30 years the model was
significantly advanced by several research groups (for review see,
e.g., Izmodenov 2000, 2002).

 The structure of the Solar Wind - LIC interaction region is shown in Figure 1.
Contact discontinuity, or {\it the heliopause} (HP), separates the
solar wind and interstellar plasmas. The heliopause is an obstacle
for both the supersonic (the Mach number is about 10) solar wind,
and the supersonic (the Mach number is about 2) interstellar gas.
A shock has to be formed in the case of supersonic flow around an
obstacle. The supersonic solar wind passes through {\it the
termination shock} (TS) to become subsonic. {\it The bow shock}
decelerates the local interstellar gas from supersonic to
subsonic. The whole region of the solar wind interaction is called
{\it the heliospheric interface}.

Note, however, that in the case when effect of interstellar atoms
is not taken into account, the qualitative picture of the tail
flow pattern is more complex. The solar wind flow is subsonic in
the nose part of the region between the termination shock and the
heliopause. Then the flow passes through the sonic line (Baranov
and Malama, 1993) and becomes supersonic. This results in
formation of the Mach disk (MD), tangential discontinuity (TD) and
reflected shock (RS) in the tail region (figure 1a).

The interstellar H atoms interact with plasma component by charge
exchange and strongly influence locations of the shocks and the
heliopause and the structure of the heliospheric interface. The
main difficulty to model the heliospheric interface is a large
mean free path of the H atoms with respect to charge exchange. The
mean free path is comparable with the characteristic size of the
heliosphere. Therefore, to describe interstellar H atom flow in
the heliospheric interface it is necessary to solve kinetic
equation. A self-consistent two-component (plasma and H atoms)
model of the heliospheric interface was proposed by Baranov et al.
(1991) and realized by Baranov and Malama (1993). The latter paper
also presented the first numerical simulations of the heliotail.
Figure 1 compares the geometrical pattern - locations of the two
shocks and the heliopause - for the model, which takes into
account the influence of interstellar H atoms, with the model,
which do not take into account the neutral component. It is seen
that the discontinuities are significantly closer to the Sun in
the case with atoms. In the heliotail region the structure of the
plasma flow changes qualitatively. The termination shock becomes
more spherical and Mach Disk (MD), reflected shock (RS) and
tangential discontinuity (TD) disappear (Figure 1).

The model of the heliospheric interface allows answering two
fundamental questions: 1. Where is the edge of the solar system?
2. How far is the influence of the solar system on the surrounding
interstellar medium?

To give an answer on the first question we need to define the
solar system boundary. It is naturally to assume the boundary is
the heliopause, which separates the solar wind and interstellar
plasmas. Note, that the influence of the solar system on the
interstellar medium is significantly far than the heliopause.
Secondary interstellar atoms, which are a result of charge
exchange of the original interstellar H atoms and solar wind
protons, disturb the interstellar gas upwind the bow shock. Detail
studies of mutual influences of charge and neutral components in
the heliospheric interface were done in Baranov and Malama
(1993,1995, 1996), Baranov et al. (1998), Izmodenov et al. (1999,
2000, 2001). However, these papers were mainly focused on the
upwind region. At the same time the study of the heliotail region
has also significant interest. In the heliotail we cannot assume
the heliopause to be the heliospheric boundary. It is seen in
Figure 1, the heliopause is not closed surface and the solar wind
fills the whole space into the downwind direction.

The goal of this work is to study the structure of the tail region
of the heliospheric interface. We focus on the effects of the
charge exchange process on the tail region.

\section{Model}

To study the effect of charge exchange on the structure of the
heliotail we used kinetic-gas dynamic model by Baranov and Malama
(1993). In this model the solar wind at the Earth's orbit is
assumed to be spherical symmetrical and not varying with time. The
interstellar flow is also assumed to be constant parallel flow.
Under these conditions the flow in the interaction region is
axisymmetric.

 To describe the charged component (electrons and protons)
we solve hydrodynamic Euler equations, where the effect of charge
exchange is taken into account in the right parts of these
equations. To calculate the flow of interstellar H atoms in the
heliospheric interface we solve kinetic equation:

\begin{eqnarray}\label{eqBoltz}
{\bf w}_{\rm H} \cdot \frac{\partial f_{\rm H}({\bf r}, {\bf
w}_{\rm H}) }{\partial {\bf r} } + \frac{{\bf F}}{m_{\rm H}} \cdot
\frac{\partial f_{\rm H} ({\bf r},{\bf w}_{\rm H}) }{\partial {\bf
w}_{\rm H} }  \\ \nonumber = - f_{\rm H} ( {\bf r}, {\bf w}_{\rm
H} ) \int | {\bf w}_{\rm H} - {\bf w}_p | \sigma^{\rm HP}_{ex} f_p
({\bf r}, {\bf w}_p) d {\bf w}_p \\ \nonumber + f_p ({\bf r}, {\bf
w}_{\rm H}) \int | {\bf w}_{\rm H}^* - {\bf w}_{\rm H} |
\sigma^{\rm HP}_{ex} f_{\rm H} ({\bf r}, {\bf w}_{\rm H}^* ) d
{\bf w}_{\rm H}^* \\ \nonumber- ( \beta_i + \beta_{\rm impact} )
f_{\rm H} ( {\bf r}, {\bf w}_{\rm H} ).
\end{eqnarray}

Here $ f_{\rm H} ({\bf r}, {\bf w}_{\rm H}) $ is velocity
distribution function of H atoms; $ f_p( {\bf r},{\bf w}_p ) $ is
locally Maxwellian velocity distribution of protons; $ {\bf w}_p $
and $ {\bf w}_{\rm H}$ is individual velocities of protons and H
atoms, respectively; $ \sigma^{\rm HP}_{ex} $ is the cross section
of the charge exchange of H atoms and protons; $\beta_i $ is the
photoionization rate; $ m_{\rm H}$ is the mass of H atom; $
\beta_{\rm impact} $ is the electron impact ionization; and $ {\bf
F} $ is the sum of the solar gravitation and radiation pressure
forces.

The main process of the plasma-neutral coupling is the charge
exchange process $H + H^+ \rightarrow H^+ + H$. Photoionization
and electron impact ionization are also taken into account in
equation ($\ref{eqBoltz}$). The interaction of charged and neutral
components results in exchange of mass, momentum and energy
between these components. The source term $Q=(q_1, q{2,z},
q_{2,r}, q_3)^T$ is in the right part Euler equation for charged
component. Here $q_1$, $\vec{q}_2={(q_{2,z},q_{2,r})^T},$ $q_3$
are sources of mass, momentum and energy, respectively. The source
terms are integrals of H atom velocity distribution function
$f_H$:
\[
q_1= n_{\rm H} \cdot (\beta_i + \beta_{impact}), n_{\rm H}= \int
f_{\rm H}({\bf w}_{\rm H}) d {\bf w}_{\rm H},
\]
\begin{eqnarray}
\vec{q}_2 =  \int (\beta_i+\beta_{impact}) {\bf w}_{\rm H} f_{\rm
H}({\bf w}_{\rm H}) d {\bf w_{\rm H}} + \\ \nonumber
 \int \int u \sigma_{ex}^{HP}(u) ({\bf w}_{\rm H} - {\bf w}_p)
f_{\rm H}({\bf w}_{\rm H}) f_p({\bf w}_p) d {\bf w_{\rm H}} d {\bf
w}_p,  \nonumber
\end{eqnarray}
\begin{eqnarray}
q_3 =  \int (\beta_i+\beta_{impact}) \frac{{\bf w}_{\rm H}^2}{2}
f_{\rm H}({\bf w}_{\rm H}) d {\bf w}_{\rm H} + \\ \nonumber
  \frac{1}{2} \int \int u \sigma_{ex}(u) ({\bf w}_{\rm H}^2 - {\bf w}_p^2)
f_{\rm H}({\bf w}_{\rm H}) f_p({\bf w}_p ) d {\bf w}_{\rm H} d
{\bf w}_p.  \nonumber
\end{eqnarray}
Here $u=|{\bf w}_{\rm H} - {\bf w}_p|$ is relative atom-proton
velocity.

 For boundary
conditions in the unperturbed LIC we adapt the velocity $V_{LIC} =
25$ km/s, interstellar H atom and proton number densities are 0.2
cm$^{-3}$ and 0.07 cm$^{-3}$. The temperature of the LIC was
assumed ~6000 K. Velocity, number density and Mach number at the
Earth orbit are 450 km/s, 7 cm$^{-3}$ and 10, respectively.
Velocity distribution function of H atoms is assumed to be
Maxwellian.

Euler equations with the source term $\vec{Q}$ were solved
self-consistently with the kinetic equation for H atoms. To get
the self-consistent solution we used the iterative method. The
kinetic equation was solved by Monte-Carlo method with splitting
of trajectories. Unlike the previously published papers based on
Baranov-Malama model, we performed the calculations in extended
computation region toward the heliotail. We performed computations
up to 50000 AU along the axis of symmetry and up to 5000 AU in the
direction perpendicular to the axis of symmetry. To estimate
divergence of chosen numerical scheme we used different
computational grids. Dependence of the numerical solution from
outer boundary conditions was estimated by variation of the
computational domain in the tail region.

\begin{figure}
\noindent\includegraphics[width=\hsize]{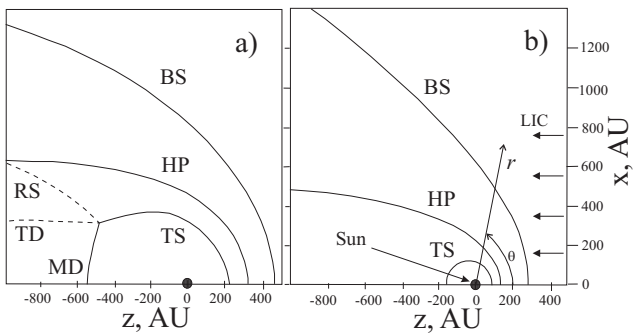}
\caption{ The structure of the  heliospheric interface is the
region of the solar wind interaction with the interstellar medium.
HP is the heliopause, TS is the heliospheric termination shock, BS
is the bow shock, MD is the Mach disk, TD is the tangential
discontinuity, RS is the reflected shock. The right figure
corresponds to the case with no H atoms, while the left figure
corresponds to the solution, when interstellar H atoms are taken
into account. \label{fig1}}
\end{figure}

\section{Qualitative analysis}

In this work we consider the effect of the charge exchange
processes ($H + H^+ \rightarrow H^+ + H$) on the plasma flow in
the tail region of the heliospheric interface. The supersonic
solar wind passes through the heliospheric termination shock,
where its kinetic energy transfers into the thermal energy. Let us
assume now that in the tail region the surface of the heliopause
is parallel to the direction of the interstellar flow. This is in
accordance with our numerical simulations. Under this assumption
the solar wind can be considered as a flow in a nozzle having
constant cross-section. Our computations show that in the case
with no H atoms the solar wind pressure downstream the termination
shock is several times smaller than the interstellar pressure.
Under these conditions the solar wind flow decelerates and has
some minimal value at infinity. The minimal value is determined by
the parameters of the solar wind downstream the termination shock
and the interstellar pressure. Neither the interstellar proton
number density nor the relative Sun-LISM velocity does not
determine the minimal velocity. Therefore, in the case with no
atoms in the frame of hydrodynamic approach it is possible to find
solution where the solar wind (and, therefore, the solar system)
extended in the heliotail up to infinity. Such a qualitative
consideration can be easily generalized, when the heliopause is
not parallel to the axis of symmetry. Then the solar wind flow can
be considered as a flow in convergent or expanding nozzle.

Qualitatively other situation is realized in the case when
interstellar H atoms are taken into account. Our calculations show
that in this case the solar wind pressure downwind the termination
shock is larger than the interstellar pressure. The solar wind
should be accelerated in this case by the pressure gradient.
However, interstellar atoms play significant role due to charge
exchange. The interstellar atoms fulfill the heliotail due to
their large mean free path. Among the heliospheric H atoms the
part of original (or primary) interstellar atoms increases
significantly with the heliocentric distance. The temperature
(6000 K) and velocity (25 km/s) of the primary interstellar atoms
are smaller than the velocity (100 km/s) and temperature (100000
K) of the post shocked solar wind. New protons, which are born
from the interstellar H atoms, have smaller average and thermal
velocities than original solar protons. Therefore, the charge
exchange process leads to effective cooling and deceleration of
the solar wind. The acceleration of the solar wind by the pressure
gradient with one side and the deceleration of the solar wind by
the charge exchange process may result in the heliopause is not
always parallel to the axis of symmetry. Since the part of primary
interstellar atoms increases with the increasing of the
heliocentric distance, it is naturally to expect approach of the
solar wind velocity, density and temperature to their interstellar
values.

Despite a number of assumptions, qualitative analysis given in
this section is confirmed by our numerical calculations. In the
next section we present and discuss results of numerical
calculations.

\begin{figure}
\noindent\includegraphics[width=\hsize]{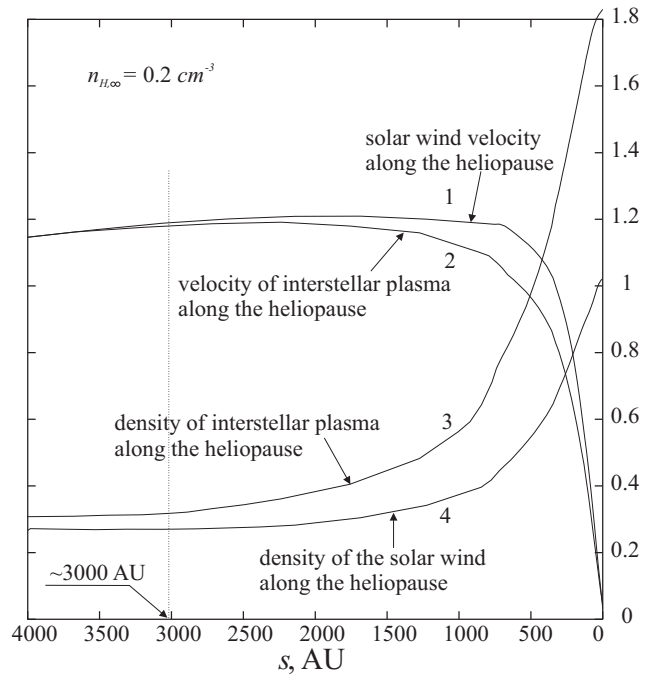}
\caption{ Velocities (curves 1 and 2) and densities (curves 3 and
4) from both sides of the heliopause as a function of the
heliocentric distance along the heliopause. Curves 2 and 3
correspond to interstellar side; curves 1 and 4 correspond to
solar wind side. The velocities and densities are normalized to
their interstellar values. \label{fig2}}
\end{figure}
\begin{figure}
\noindent\includegraphics[width=\hsize]{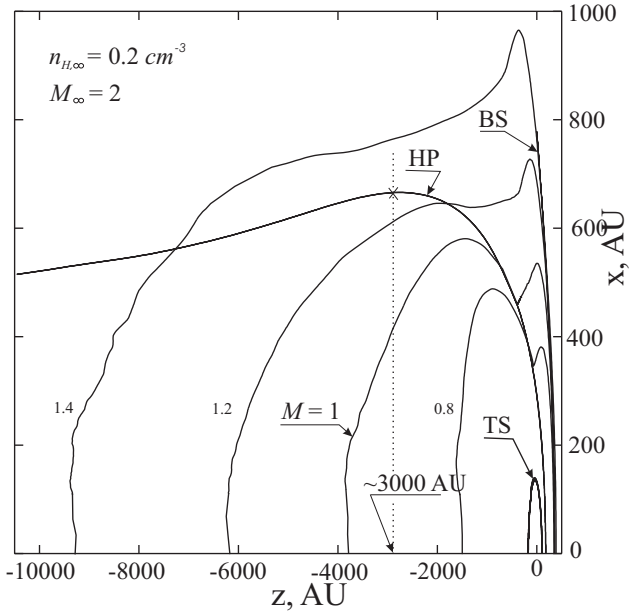}
\caption{ Isolines of Mach number M. It is seen that on the
distances more than 4000 AU into the heliotail direction the solar
wind flow is supersonic. The Mach number increases with increase
of the heliocentric distance and approaches its interstellar
number. \label{fig3}}
\end{figure}

\begin{figure}
\noindent\includegraphics[width=\hsize]{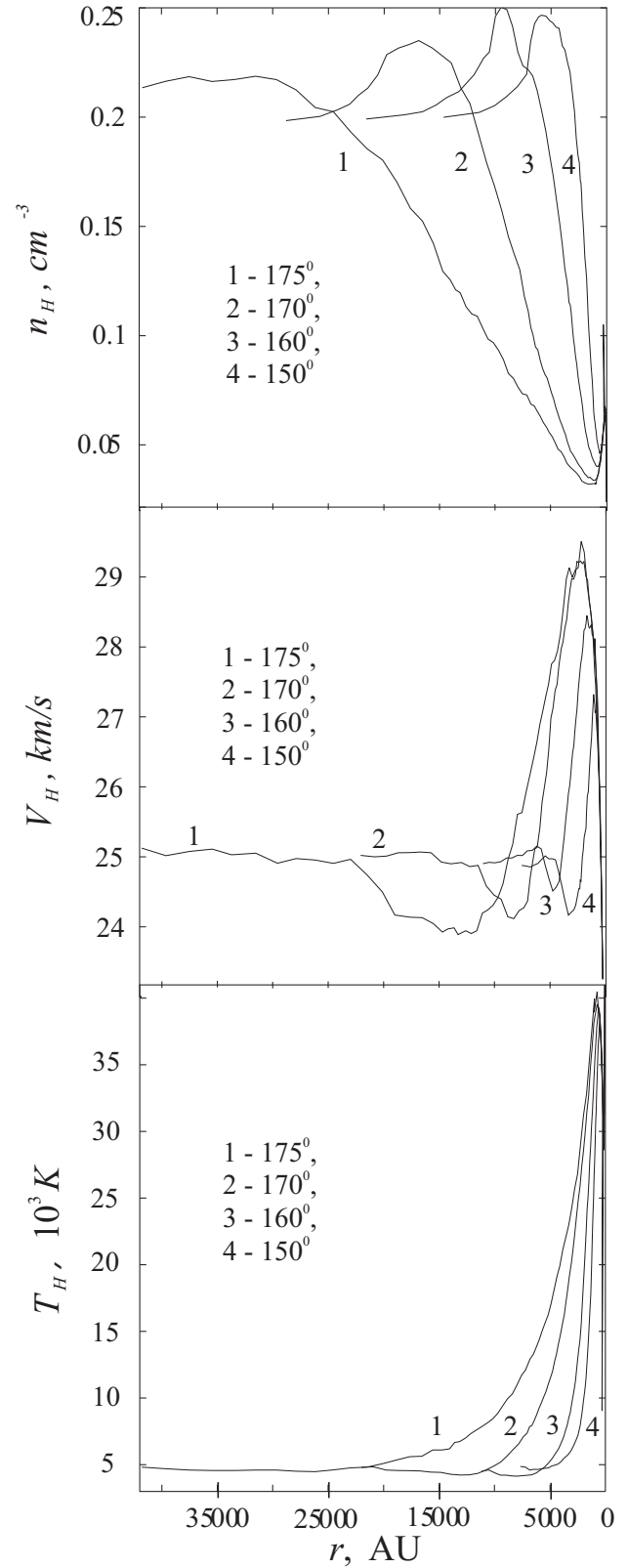}
\caption{ Number density, velocity and temperature of the
interstellar H atom along downwind lines of sights  $\theta$ =
150, 160, 170, 175 degrees. \label{fig4}}
\end{figure}

\section{Results and Discussion}

Results of our calculation confirm the qualitative analysis given
above. Distributions of plasma parameters in the heliotail region
are presented in figures 2 and 3. Figure 2 presents the
distributions of density and velocity of plasma along the
heliopause. The parameters are shown from both interstellar and
solar wind sides of the heliopause. In the classical hydrodynamics
two conditions determine the tangential discontinuity. The
conditions are 1) no mass transport through the discontinuity, 2)
balance of pressures on the both sides of the discontinuity. These
conditions permit a jump of density and tangential velocity
through the heliopause. In the case with presence of interstellar
H atoms the jump of density and pressure becomes weaker with
increase of the distance calculated along the heliopause from its
nose. This is due to mass transport caused by charge exchange. For
$z \approx -3000$ AU, where z is the distance along the axis of
symmetry and sign "-" means the direction toward the heliotail,
the jump of density and tangential velocity disappears (Figure 2).

The velocity of the solar wind is about 100 km/s downstream the
termination shock. Then the velocity becomes smaller due to new
injected by charge exchange protons and approaches the value of
interstellar velocity. The solar wind becomes also cooler due to
charge exchange. The interstellar Mach number is $\sim$2. It is
interesting to see whether the solar wind becomes supersonic due
to effective cooling. Figure 3 shows isolines of Mach numbers in
the heliospheric interface. The solar wind passes through the
sound velocity at about $z \approx$-4000 AU, then the Mach number
increases approaching its interstellar value. The heliopause is
also shown in Figure 3. The line $z=-3000 $ AU shows the distance
where when jump of density and velocity through the heliopause
disappears.

\begin{figure}
\includegraphics[width=\hsize]{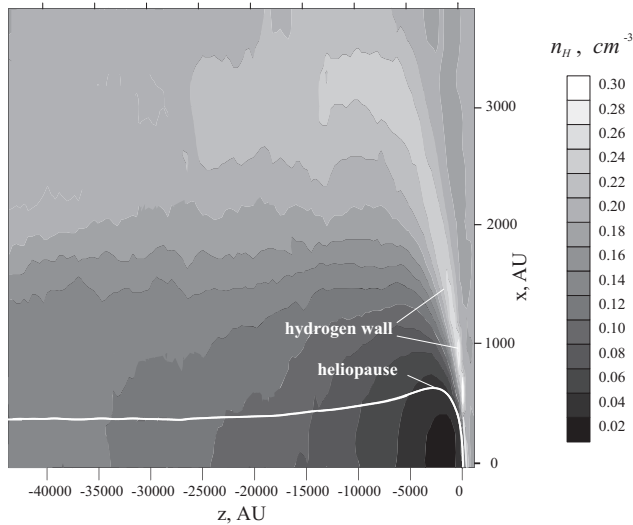}
\caption{ 2D distribution of H atom number density in the
heliospheric interface. At heliocentric distances $\sim$ 400000 AU
number density of H atoms is closer to the interstellar value. The
hydrogen wall, the in crease of H atom number density in front of
the heliopause, is also shown. The intensity of the hydrogen wall
falls with the increase of the heliocentric distance.}
\end{figure}

Figures 4 and 5 show the distribution of the interstellar atoms in
the tail part of the heliospheric interface. Figure 4 represents
the densities, velocities and temperatures of the interstellar
hydrogen along the different downwind directions. The angle
$\theta$ in Figure 4 is the angle between line-of-sight and upwind
directions (Figure 1). Parameters of H atoms approach their
interstellar values on distances less than 20000 AU for all
line-of-sights. The approach is faster for smaller $\theta$. It is
interesting to note, that the hydrogen wall, the increase of H
atom number density in the region between the heliopause and the
bow shock [Baranov et al, 1991; Izmodenov, 2000) is visible even
for large $\theta \approx 150-170^o$. Two dimensional distribution
of H atom number density in the tail region is shown in Figure 5.

It is important that to get numerical solution of the heliotail
plasma becomes possible due to the charge exchange process, which
makes the solar wind to be supersonic at the outer boundary. This
allows fulfilling correct boundary conditions.

In this work we considered influence of the charge exchange
process only. In future, influences of different hydrodynamic and
plasma instabilities, interstellar and heliospheric magnetic
fields on the heliotail structure must to be considered. The
processes of reconnection can be also important.

%%%%%%%%%%%%%%%%%%%%%%%%%%%%%%%%%%%%%%%%%%%%%%%%
%% BACKMATTER
%%%%%%%%%%%%%%%%%%%%%%%%%%%%%%%%%%%%%%%%%%%%%%%%
\section{Summary}

In this paper we consider effects of charge exchange on the
structure of the heliotail region. In particular, it was shown
that

1. The charge exchange process change the solar wind -
interstellar interaction flow qualitatively in the tail region.
The termination shock becomes more spherical and Mach disk,
reflected shock and tangential discontinuity disappear (Figure 1).
The jumps of density and tangential velocity through the
heliopause become smaller into the heliotail and disappear at
about 3000 AU.

2. Parameters of solar wind plasma and interstellar H atoms
approach their interstellar values at large heliocentric
distances. This allows to estimate the influence of the solar
wind, and, therefore, the solar system size into the downwind
direction as about 20000- 40000 AU. Unlike the upwind direction
the solar system boundary has diffusive nature in the heliotail.

3. The supersonic character of the solar wind flow in the
heliotail allows us to perform correct numerical calculations,
which are not possible in the case without H atoms.

\bigskip
We thank A. Myasnikov for his gas dynamical code and Yu. Malama
for his Monte-Carlo code. We also thank to Baranov, Chalov, Malama
and Myasnikov for fruitful discussions. This work was supported in
part by INTAS Awards 2001-0270, RFBR grants 01-02-17551,
02-02-06011, 02-02-06012, 01-01-00759, CRDF Award RP1-2248, and
International Space Science Institute in Bern, Switzerland.

%\newpage
%\begin{thebibliography}{99}
%\begin{verse}
\bigskip
\centerline{\bf References}
\smallskip
Baranov V.B., Malama Yu.G.,
%J. Geophys. Res., V. 98, No A9, P. 15,157, (1993).
J. Geophys. Res., {\bf 98}, {\bf A9}, 15157 (1993).

%\bibitem{2}
Baranov V.B., Malama Yu.G.,
%J. Geophys. Res., V. 100, P.  14,755, (1995).
J. Geophys. Res., {\bf 100}, 14755 (1995).

%\bibitem{3}
Baranov V.B., Malama Yu.G.,
%Space  Science  Rev., V. 78, P. 305, (1996).
Space  Science  Rev., {\bf 78}, 305 (1996).

%\bibitem{4}
Baranov, V.~B., Krasnobaev, K.~V., Kulikovsky, A.~G.,
%Докл. АН СССР, т. 194, с. 41, (1970).
Sov. Acad. Reports, {\bf 194},  41 (1970), in Russian.

%\bibitem{5}
Baranov V.B., Lebedev M.G., Malama Yu.G.,
%Astrophys. J., V. 375, P. 347, (1991).
Astrophys. J., {\bf 375}, 347 (1991).

%\bibitem{6}
Baranov V.B., Izmodenov V.V., Malama Yu.G.,
%J. Geophys. Res., V. 103, No A5, P. 9575, (1998).
J. Geophys. Res., {\bf 103}, {\bf A5}, 9575 (1998).

%\bibitem{7}
Baranov V.B., Lebedev M.G., Malama Yu.G.,
%Astrophys. J., V. 375, P. 347, (1991).
Astrophys. J., {\bf 375}, 347 (1991).

%\bibitem{8}
Witte M., Banaszkiewicz M., Rosenbauer H., Recent Results on the
Parameters of the Interstellar Helium from the Ulysses/Gas
Experiment, Space Science Rev.,
%V. 78, Issue 1/2, P. 289-296, (1996).
{\bf 78}, issue 1/2, 289 (1996).

%\bibitem{9}
Izmodenov V.V., Astrophys. Space Sci., {\bf 274}, issue 1/2, 55
(2000).

%\bibitem{10}
Izmodenov V.V.,
%COSPAR Colloquia Series, {\bf 11}, 23 (2000).
Proceeding of Special COSPAR Colloquium In Honour of Stanislaw
Grzedzielski, Leaving Executive Director of COSPAR, COSPAR
Colloquia Series, in press (2002).

%\bibitem{11}
Izmodenov V.V., Lallement R., Malama Yu. G.,
%Astron. Astrophys., V. 342, P. L13, (1999).
Astron. Astrophys., {\bf 342}, L13 (1999).

%\bibitem{12}
Izmodenov V.V., Gruntman M., Malama Yu.G.,
%J. Geophys. Res., V. 106 , No. A6 , P. 10,681-10,690, (2001).
J. Geophys. Res., {\bf 106}, {\bf A6}, 10681 (2001).

%\bibitem{13}
Lallement R.,
%Space Science Rev., V. 78, P. 361, (1996).
Space Science Rev., {\bf 78}, 361 (1996).

%\bibitem{14}
Malama Yu.G.,
%Astrophys. Space Sci., V. 176, P. 21, (1991).
Astrophys. Space Sci., {\bf 176}, 21 (1991).

\vspace{3cm}

\end{document}